\documentclass[%
 aps,
 prl,
 amsmath,amssymb,
reprint,%
]{revtex4-2}

\usepackage{graphicx}
\usepackage{dcolumn}
\usepackage{bm}

\usepackage[utf8]{inputenc}
\usepackage[T1]{fontenc}
\usepackage{mathptmx}
\usepackage{etoolbox}
\usepackage{siunitx}
\usepackage{xcolor}
\usepackage{placeins}
\usepackage{textgreek}
\usepackage{amsmath}
\usepackage[colorlinks=false]{hyperref}
\usepackage{cleveref}
\usepackage{dcolumn}
\makeatletter
\def\@email#1#2{%
 \endgroup
 \patchcmd{\titleblock@produce}
  {\frontmatter@RRAPformat}
  {\frontmatter@RRAPformat{\produce@RRAP{*#1\href{mailto:#2}{#2}}}\frontmatter@RRAPformat}
  {}{}
}%
\makeatother
\begin{document}

\preprint{APS/123-QED}

\title{Direct Measurement of Electron Heating in Electron-Only Reconnection in a Laboratory Mini-Magnetosphere}

\author{Lucas Rovige$^{1}$, Filipe D. Cruz$^{2}$, Timothy Van Hoomissen$^{1}$,
Robert S. Dorst$^{1,3}$, Carmen G. Constantin$^{1}$, Stephen Vincena$^{1}$, Luis O. Silva$^{2}$, Christoph Niemann$^{1}$, and Derek B. Schaeffer$^{1}$}

\affiliation{$^{1}$University of California--Los Angeles, Los Angeles, CA 90095, USA}
\affiliation{$^{2}$GoLP/Instituto de Plasmas e Fusão Nuclear, Instituto Superior Técnico, Universidade de Lisboa, 1049-001 Lisboa, Portugal}
\affiliation{$^{3}$Lawrence Livermore National Laboratory, 7000 East Avenue, Livermore, California 94550, USA}

\date{\today}

\begin{abstract}

We report on the experimental observation of electron heating in electron-only magnetic reconnection in laser-driven laboratory mini-magnetospheres on the Large Plasma Device (LAPD) at the University of
California, Los Angeles. In this experiment, a fast-flowing plasma impacts a pulsed magnetic dipole embedded within LAPD's magnetized ambient plasma, creating an ion-scale magnetosphere and driving electron-only magnetic reconnection between the background and dipole field lines. The electron velocity distribution is measured across the reconnection region using non-collective Thomson scattering, enabling determination of electron temperature and density. Significant electron heating is observed in the electron diffusion region, increasing from an initial temperature of 1.8 eV to 9.5 eV, corresponding to a 40\% conversion of Poynting flux into electron enthalpy flux. Particle-in-cell  simulations that provide insights into the heating mechanisms are also presented.
\end{abstract}

\maketitle

Magnetic reconnection \cite{yamada_magnetic_2010} is a fundamental plasma process where the breaking and topological rearrangement of magnetic field lines abruptly releases stored magnetic energy to the plasma and drives explosive events in many different environments across the universe, from planetary magnetospheres, pulsars, and solar flares to disruptions in fusion devices and other laboratory experiments. Magnetic reconnection can energize the ions and electrons of the plasma through both directed kinetic energy and thermal heating and understanding the partition of this energy transfer as well as the mechanisms behind it is necessary to understand the role of reconnection in particle acceleration, heating, and large-scale energy transport in space and astrophysical systems. \par
Recently, enabled by the high-resolution measurements of the Magnetospheric Multiscale (MMS) Mission  \cite{Burch16,Wilder17}, a new regime of magnetic reconnection has been identified in which only electrons participate in the reconnection process while the ion population remains largely decoupled from the dynamics \cite{phan_electron_2018,sharma_pyakurel_transition_2019}. This electron-only reconnection has been observed in regions of Earth's magnetosphere \cite{phan_electron_2018,man_observations_2020}, in laboratory experiments \cite{shi_laboratory_2022,shi_using_2023,Chien2023,rovige_laboratory_2024} and has also attracted interest as its dynamics may be relevant as the onset of classical ion-coupled reconnection \cite{hubbert_electron-only_2022,lu_electron-only_2022}.  
In recent work, Lunar magnetic reconnection from mini-magnetospheres formed by the interaction of the solar wind with ion-scale crustal magnetic anomalies has been observed \cite{sawyer_does_2023}, while the demagnetization of the ions and the small, ion-scale size of the system suggested this reconnection was electron-only. This \textit{in-situ} observation has been confirmed by subsequent numerical \cite{stanier_intermittent_2024} and experimental \cite{rovige_laboratory_2024} results reproducing mini-magnetospheres and observing electron-only reconnection.  Additionally, the MAVEN mission enabled the detection of many reconnection events in martian mini-magnetospheres \cite{harada_survey_2017,harada_magnetic_2018,wang_maven_2021,qiu_observations_2024}, a majority of which were ion-coupled, but principally because the instruments on MAVEN cannot resolve the electron scale. Nevertheless, it is expected that a significant part of the reconnection events on Mars are electron-only \cite{harada_survey_2017,wang_magnetic_2023}, therefore laboratory experiments of reconnection in mini-magnetospheres would provide valuable insight in electron-only reconnection in a context relevant to the Moon, Mars and other small-scale magnetospheres such as near Mercury or Ganymede \cite{slavin_messenger_2007,ebert_evidence_2022}.\par
Previous laboratory studies have focused on understanding the partition of energy transfer in reconnection. In classical ion-coupled reconnection, \textit{Yamada et al.} \cite{yamada_conversion_2014} observed that 14\% of the inflowing Poynting Flux was converted to electron enthalpy flux, while \textit{Shi et al.} \cite{shi_laboratory_2022} showed that electron-only reconnection is much more efficient at transferring energy to the electrons, as they measured a 70\% transfer of Poynting flux towards electron enthalpy. In their experiment, electron-only reconnection was driven by the merging of two symmetric flux tubes, in the presence of a strong guide field, and a subsequent experiment measuring anisotropy \cite{shi_using_2023} showed that the electron heating was dominated by the energization from parallel electric fields.\par

In this Letter, we report the first direct measurement of electron heating during electron-only magnetic reconnection in a laboratory magnetosphere, using non-collective Thomson scattering to obtain spatially and temporally resolved profiles of electron temperature and velocity distribution. This work provides new insights into how reconnection fundamentally transforms magnetic energy into particle energy in an asymmetric, anti-parallel (no guide-field) configuration, probing key physics of energy conversion in collisionless plasmas. Beyond its fundamental importance, this dynamic is directly relevant to space environments such as the Moon and Mars, where reconnection at ion-scale magnetic anomalies likely occurs on the electron scale. Because \textit{in situ} spacecraft measurements remain sparse and challenging in these regions, our laboratory approach provides critical access to understanding how such systems evolve. We notably observe a fourfold temperature increase in the diffusion region, as well as a significant conversion from the inflowing Poynting flux to the electron enthalpy flux that is estimated to be 40\%. Particle-in-Cell (PIC) simulations provide more insight into the mechanism leading to electron heating, and show that both the energization from the out-of-plane reconnection electric field, and a Fermi-like reflection energization from outflowing reconnected field lines contribute to electron heating. \par

\begin{figure}[ht!]
\includegraphics[width=1\linewidth]{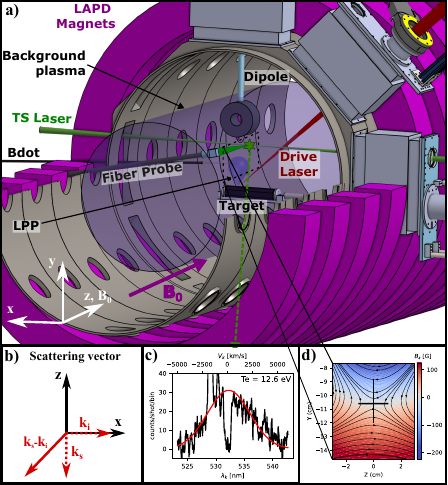}
\caption{\label{fig:setup} a) Schematic of the experiment on the LAPD. b) Scattering vector  $\mathbf{k_s}- \mathbf{k_i}$.  c) Example of a Thomson scattering spectrum obtained by averaging over 400 shots (black) and its gaussian fit used to retrieve density and temperature (red). d) Initial magnetic field in the z-direction $B_z$ and magnetic field lines in the reconnection (Y-Z) plane, measured with a B-dot probe. The centered vertical and horizontal lines represent the TS measurement regions.    }
\end{figure}

\textit{Experimental setup and Thomson scattering measurements.} The experiment is schematized in Fig.\ref{fig:setup}a. A high-repetition rate (1 Hz), 20~ns laser with 10~J of energy per pulse is focused on a graphite target and drives a fast moving carbon plasma \cite{Niemann2013,Schaeffer2016}
that expands into the field generated by a pulsed magnetic dipole embedded into the hydrogen background discharge plasma and axial magnetic field ($\mathbf{B_0} = B_0 \mathbf{\,\hat z}$, with $B_0=210$ G) generated by the Large Plasma Device (LAPD) \cite{Gekelman2016}. The laser-plasma expands at a velocity $V_0 = 250$~km/s, corresponding to a super-Alfvenic flow with $M_A = 1.7$. The field of the dipole oriented to be anti-parallel to the background field on the outside, so that a magnetic null point is formed in the region between the dipole and the target, similar to our previous experiment \cite{rovige_laboratory_2024}.  The time-resolved magnetic field is measured with a magnetic flux (B-dot) probe \cite{Everson2009} that is moved between shots to map out the field geometry. \par
The electron velocity distribution is measured using a motorized non-collective Thomson scattering (TS) diagnostics \cite{ghazaryan_thomson_2022,zhang_two-dimentional_2023} that can scan across a $2\times2$~cm$^2$ region in (y,z) around the reconnection point. A 0.5~J, 4~ns laser of wavelength $\lambda_{T}=532$~nm is focused along the x-axis in the region of interest, and the scattered light is collected along the z-axis by a fiber-probe and sent to a spectrometer. The fiber probe consists of 20 individual fiber cores aligned along the $x$-direction to enhance the collected signal. The effective scattering volume defined by the intersection of the Thomson scattering laser beam and the projected image of the fiber bundle through an imaging lens is approximately $15 \times 0.1 \times 0.1~\text{mm}^3$ in the $(x, y, z)$ directions. As a result, the temperature measurements are well resolved in the $(y, z)$ plane but averaged over $15~\text{mm}$ along the $x$-direction. \par
Figure \ref{fig:setup}c shows an example of a TS spectrum and associated velocity distribution, obtained in the laser-plasma. The drop in signal level at the center of the distribution is caused by a physical notch of 1.5~nm width in the spectrometer blocking stray light from the laser at 532~nm. Two emission lines from neutral carbon at 529~nm and 530~nm are visible in some spectra and removed from the fitting procedure. The electron temperature is estimated by calculating the width of a gaussian fit of the electron velocity distribution, the relative density is retrieved from the area under the fitted curve, and cross-calibrated with the background density measurement from the interferometer and Langmuir probe.  The electron flow-velocity in the scattering direction $V_k$ is obtained by measuring the Doppler-shift of the peak of the distribution. The background electron temperature is measured to be $T_{e0}=1.8\pm0.4$~eV. Since the incident vector $\mathbf{k_i}$ is purely along $x$ and the scattering vector $\mathbf{k_s}$ purely along $z$, the TS measurement vector $\mathbf{k} = \mathbf{k_s}- \mathbf{k_i}$ is at 45$^\circ$ from both $x$ and $z$ axes (see Fig. \ref{fig:setup}b). \par

\begin{figure}[ht!]
\includegraphics[width=1\linewidth]{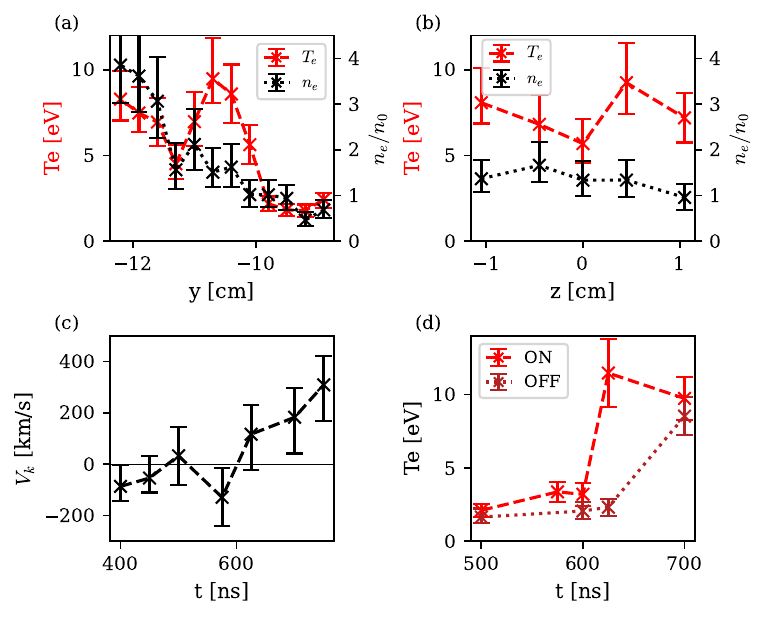}
\caption{\label{fig:TS} (a)-(b) Electron temperature and density measured by Thomson scattering along the y and z lineouts across the reconnection zone as represented on Fig.~\ref{fig:setup}c at $t = 650$~ns. Each position data point is obtained from a spectrum averaged over 400~shots. (c) Electron velocity along $\mathbf{k}$, $V_k$ at $y=-10.5$~cm and $z= -1.5$~cm (outflow) for different times, measured from the Doppler shift of the TS spectra. (d) Electron temperature at the reconnection point $y=-10.5$~cm and $z = 0$ with the dipole turned ON and OFF, for different times of the experiment.}
\end{figure}

Figures~\ref{fig:TS}a and \ref{fig:TS}b show the electron temperature and density retrieved from Thomson scattering spectra along the $y$- and $z$-axes, respectively, across the reconnection point located at $y = -10.5$~cm, $z = 0$, at $t = 650$~ns, a time at which magnetic field lines are actively reconnecting. The vertical profile reveals localized electron heating centered around the reconnection point, with a peak temperature of $T_e = 9.5 \pm  \left( \substack{2.4 \\ 1.4} \right)$~eV and a corresponding density compression reaching $n_e = 1.5$--$2~n_0$. A second region exhibiting similar heating and a stronger, fourfold density enhancement is observed further from the dipole, for $y < -11.3$~cm, associated with the front of the laser-driven plasma. Along the \( z \)-direction, both temperature and density remain relatively uniform across the current sheet. \par

\begin{figure*}[ht!]
\includegraphics[width=1\linewidth]{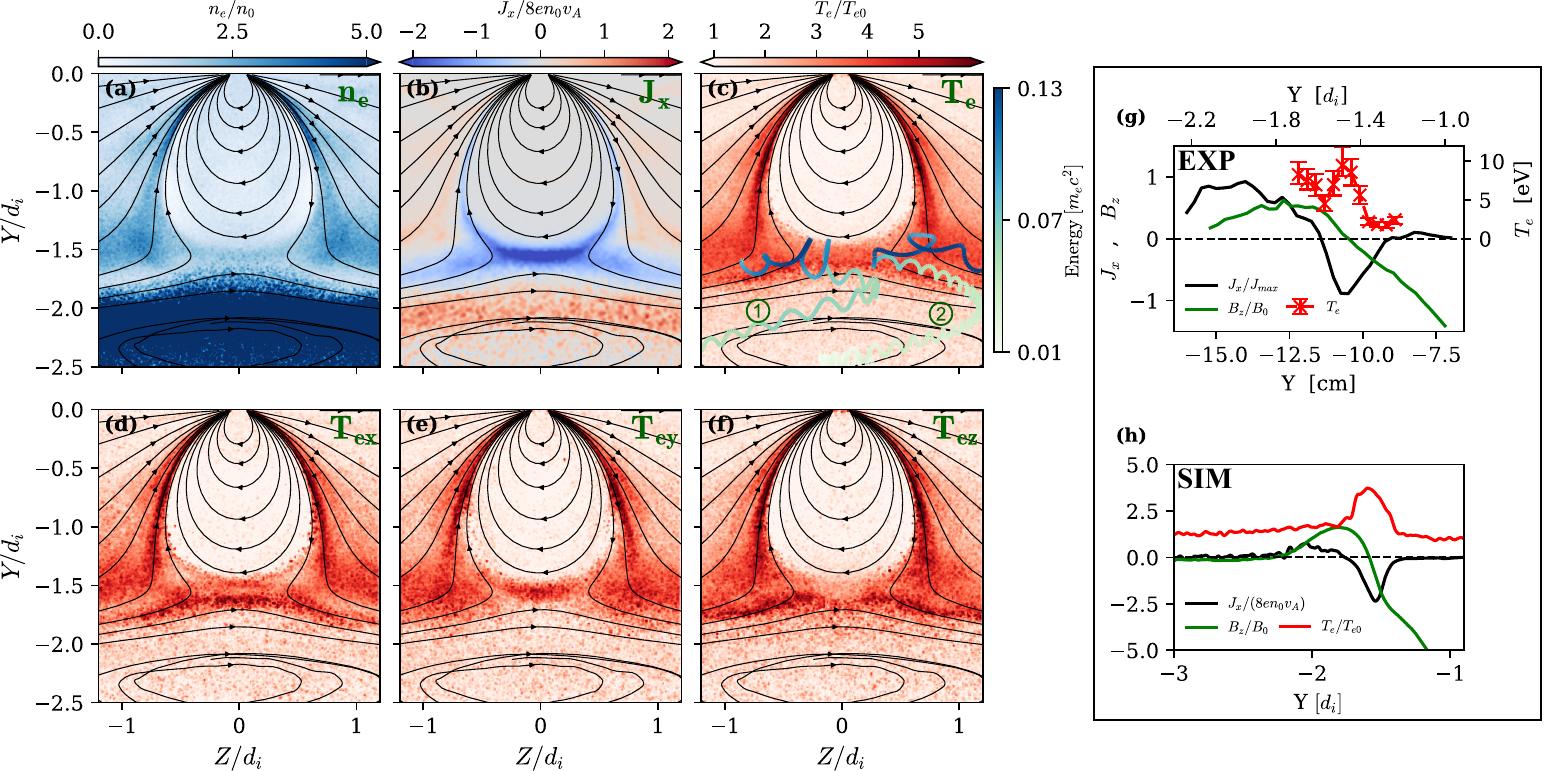}
\caption{\label{fig:sim}\textit{Simulation result at simulation time $t= 3.9~\omega_{ci}^{-1}$ and $x=0$.} The black streamlines represent the magnetic field lines and  (a) Normalized electron density (b) Normalized out of plane current density $J_x$. (c) Electron temperature normalized to the initial temperature. The trajectories of two electrons gaining a significant amount of energy in the reconnection layer are plotted, with their energy in green-blue color scale. (d)-(f) Components of the electron temperature $T_{ex},T_{ey},T_{ez}$ respectively along the $x,y,z$ directions, normalized to the initial temperature. (g)-(h) Vertical lineouts of the out of plane current $J_x$ (black), reconnection magnetic field $B_z$ (green) and electron temperature $T_e$ (red): (top) in the experiment, at $t=650$~ns and (bottom) in the simulation.}
\end{figure*}

At the reconnection point, the temperature increase is \( \Delta T_e = 7.7 \pm \left( \substack{2.4 \\ 1.4} \right)\,\text{eV}\) compared to a location 1\,cm (\( 2\delta \)) away in the \( y \)-direction. This corresponds to an enthalpy production rate $\Delta H = \gamma/(\gamma -1)\,n_ek_b\Delta T_e\,(2\delta\cdot2L)/\tau=676$~kW/m, with  the specific heat ratio $\gamma = 5/3$, the current sheet width $2\delta = 10$~mm, the current sheet length $2L = 30$~mm, and the electron transit time through the current sheet $\tau = \delta/V_0 = 28$~ns. This can be compared to the dissipated Ohmic power from the reconnection current flow, calculated using the Spitzer resistivity: $P_{Ohm} = \eta J_x^2\,(2L\cdot 2\delta) = 7$~kW/m$\sim0.01\Delta H$, indicating that the vast majority of the heating is not due to collisional dissipation, but instead arises from collisionless processes. In order to estimate the fraction of magnetic energy going towards heating the electrons, we calculate the ratio of electron enthalpy production to the incoming Poynting flux: $\frac{\gamma/(\gamma-1)n_e k_b\Delta T_e }{B^2_{rec,z}/\mu_0} = 40 \pm \left( \substack{13 \\ 7} \right)\%$, using the peak reconnecting magnetic field measured in Fig. \ref{fig:sim}g on the laser-plasma side $B_{rec,z} = 0.57\,B_0 = 120$~G.

Figure~\ref{fig:TS}c displays the electron bulk flow velocity projected along the Thomson scattering vector, $V_k$, measured at $z = -1.5$~cm and $y = -10.5$~cm, placing it in the outflow region. It shows that after 600~ns, a bulk flow along $+\mathbf{k}$ is measured, with $V_k=310\pm \left( \substack{110 \\ 140} \right)$~km/s. Since the projection vector is $\mathbf{k} = -\mathbf{z} - \mathbf{x}$ the positive value $V_k$ is most likely associated with a negative electron outflow $V_z$, since the electron flow in the x direction is expected to be towards the positive $x$ from the negative reconnection current measured, and therefore contribute negatively to $V_k$. We assume this contribution from the flow along $x$ is negligible since the measurement is performed on the very edge of the current sheet, therefore the outflow velocity $V_z$ can be estimated to be $V_z = -\sqrt{2}V_k = -440 \pm \left( \substack{200 \\ 150} \right)$~km/s. This outflow velocity is super-Alfvenic with $V_z = 3\,V_A$, but significantly lower than the electron Alfven-speed $V_z = 0.07\,V_{Ae}$. \par
Finally, Fig.~\ref{fig:TS}d shows that when the dipole, and thus reconnection, is turned off no significant heating is observed at $y=-10.5$~cm, $z = 0$ and $t=625$~ns while the temperature has increased to $T_e=11.5 \pm 2$~eV when the dipole is on, confirming this heating is due to magnetic reconnection. Later on, a higher temperature is also measured in the OFF case, associated with the propagation of the hot laser-plasma inside the measurement region.

\textit{PIC simulations and discussion.} To investigate the mechanisms leading to the heating, we carried out 3D PIC simulations reproducing the experiment using the code OSIRIS \cite{fonseca_osiris_2002}, using a reduced mass ratio $m_p/m_e =100$. In the simulations, a carbon plasma slab modeling the laser-produced plasma moves toward a dipolar field through a region of magnetized ambient hydrogen plasma, at a velocity $v_{0d}$ so that the Alfvénic Mach number is $M_A = 1.5$. \par
Figure \ref{fig:sim} shows 2D maps of different simulation quantities at a time where reconnection is occurring, and where the piston plasma is close to the reconnection point, as can be seen through the uniform slab of elevated electron density in panel (a). A strong electron scale reconnection current sheet forms at $y = -1.55~d_i$ with a width $2\delta \sim 2~d_e = d_i/5$ (see Fig.~\ref{fig:sim}b). Figure~\ref{fig:sim}c shows significant heating of the electrons to a peak temperature $T_e \sim 3.7~T_{e0}$ around the reconnection point, in a zone of width similar to the current sheet, as well as in the outflows and the wings of the magnetosphere. This represents a temperature gain $\Delta T = 3.7~T_{e0} - 1.2~T_{e0}$ from a distance of $3\delta$ in y (towards the inside of the magnetosphere), yielding a ratio of electron enthalpy production to the incoming Poynting flux: $\frac{\gamma/(\gamma-1)n_e k_b\Delta T_e }{B^2_{rec,z}/\mu_0} = 49\%$, which is comparable to what we measured experimentally. This fraction of the energy contributing to heating the electrons can be compared to previous experimental work. \textit{Yamada et al.} \cite{yamada_conversion_2014} observed 14\% of the Poynting flux going to the electron enthalpy in ion-coupled reconnection without guide field. On the other hand, \textit{ Shi et al.} \cite{shi_laboratory_2022} observed a conversion of 70\% towards electron heating in electron only reconnection with a strong guide field. 
Figures \ref{fig:sim}g-h show a comparison of vertical lineouts of reconnection magnetic field, current, and electron temperature in the experiment and in the simulation, highlighting their good agreement, with a localized heating around the magnetic null point and reconnection current sheet.\par
Comparing the different components of the electron temperature in Fig.~\ref{fig:sim}d--f reveals directional anisotropies that provide insight into the mechanisms responsible for the heating. Most of the anisotropy is localized near the reconnection point, where the temperature is dominated by $T_{ex}$. This points towards the reconnection electric field $E_x$ being responsible of energizing electrons as they enter the electron diffusion region (EDR) and become unmagnetized. In the outflow regions, the parallel electron temperature component \( T_{ez} \) dominates, reaching approximately \( T_{ez} \sim 1.3\,T_e \). Farther downstream in the exhaust, the electron temperature becomes isotropic again, with all three components contributing comparably.
We tracked electrons from the simulation that went through the reconnection region and gained significant energy. The trajectories of two of these electrons are represented over the temperature map in Fig~\ref{fig:sim}c. Trajectory \textcircled{1} is typical of an electron accelerated by the reconnection electric field $E_x$ \cite{hesse_diffusion_1999,ricci_electron_2003,egedal_large-scale_2012,le_two-stage_2016}, as it enters the diffusion region, it gets trapped in the magnetic well and meanders across the width of the current sheet thus spending more time interacting with $E_x$. Trajectory \textcircled{2} is typical of another heating mechanism in the outflows, similar to the Fermi-acceleration process \cite{drake_electron_2006,dahlin_mechanisms_2014}, where electrons travel along field lines toward the reconnection point and reflect at the line’s end. As the newly reconnected field line contracts and straightens outward at the outflow velocity, it imparts an additional kick to the electron in the $z$-direction during reflection. The significant contribution of this mechanism to the heating is an important difference from what has been recently observed on PHASMA \cite{shi_using_2023} in electron-only reconnection where the presence of a strong guide field was preventing its efficiency. 

\textit{Conclusions.} These results demonstrate an important electron heating associated with electron-only magnetic reconnection in mini-magnetospheres. We observe that 40\% of the incoming Poynting flux is converted to electron enthalpy (49\% in the simulation). PIC simulations show that electron heating is sustained by two distinct processes. First, electrons gain energy directly from the reconnection electric field inside the EDR mechanism previously identified as dominant for electron-only reconnection with a strong guide field \cite{shi_laboratory_2022,shi_using_2023}. Second, a Fermi-type acceleration operating in the outflow jets contributes significantly, which was not the case on previous experiments because the presence of a strong guide field suppressed it \cite{shi_using_2023}. Future experiments will assess the relative importance of these mechanisms in mini-magnetospheres through measurements of electron-temperature anisotropy during anti-parallel, electron-only reconnection. \par

\section*{Acknowledgments}
The experiments were performed at the UCLA Basic Plasma Science Facility (BaPSF), which is a collaborative research facility funded by the U.S. Department of Energy, Fusion Energy Sciences program, with additional support by the National Science Foundation. The authors are grateful to the staff of the BaPSF for their help in carrying out these experiments. The experiments were
supported by the NSF/DOE Partnership in Basic Plasmas Science and Engineering Award Nos. PHY-2320946 and PHY-2409284, and DOE award DE-FOA-0002982 (LaserNetUS). This work is supported by the Portuguese Science Foundation (FCT), under the Project No. 2022.02230.PTDC (X-MASER) and the PhD Fellowship Grant UI/BD/154620/2022. Simulations were performed at Deucalion (Portugal), funded by FCT Masers in Astrophysical Plasmas (MAPs) i.P project 2024.11062.cPcA.A3.

\section*{Data Availability Statement}

The data that support the findings of this study are available from the corresponding author upon reasonable request.



\bibliographystyle{apsrev4-2}
\bibliography{bib}

\end{document}